\documentclass[aps,prl,twocolumn,float]{revtex4}
\usepackage{amsmath,bm,epsfig}

\let \nn  \nonumber

\let\*\cdot
\def\<{\left\langle} \def\>{\right\rangle} \def\({\left(} \def\){\right)}
 \let\~\widetilde \let\^\widehat 

\newcommand{\B}[1]{{\bm{#1}}}
\newcommand{\C}[1]{{\mathcal{#1}}}    

\let \nn  \nonumber
\usepackage{pstricks}
\usepackage{pst-node}
\usepackage[ansinew]{inputenc}
\usepackage{amssymb,amsmath}
 \let\~\widetilde
\newcommand{\Ref}[1]{(\ref{#1})}
\newcommand{\REF}[1]{Eq.~(\ref{#1})}
\def\BSE{\begin{subequations}}\def\ESE{\end{subequations}}
\let \= \equiv

\def\a{\alpha}

\def\o{\omega}

\def\be{\begin{equation}}       \def\ba{\begin{array}}

\def\ee{\end{equation}}         \def\ea{\end{array}}

\def\bea {\begin{eqnarray}}      \def\eea {\end{eqnarray}}

\def\bean{\begin{eqnarray*}}    \def\eean{\end{eqnarray*}}

\def\RA {\ \Rightarrow\ }

\def\<{\langle} \def\({\left(}  \def\>{\rangle} \def\){\right)}

\newtheorem{exi}{Example}

\begin{document}

\title{ A Model of Intra-seasonal Oscillations in the Earth atmosphere}
\author{Elena Kartashova$^{* \dag}$ and Victor S. L'vov}
 \email{Victor.Lvov@Weizmann.ac.il, lena@risc.uni-linz.ac.at}
  \affiliation{$^*$ Department
of Chemical Physics, The Weizmann Institute of Science, Rehovot
76100, Israel \\
$^\dag$ RISC, J.Kepler University, Linz 4040, Austria }

\begin{abstract}
We suggest a way of rationalizing an intra-seasonal oscillations
(IOs) of the Earth atmospheric flow   as four meteorological
relevant triads  of interacting planetary waves, isolated from the
system of all the rest planetary waves.
 Our model is independent of the topography (mountains,
etc.) and     gives a natural explanation of IOs both in the North
and South Hemispheres.  Spherical planetary waves are an example of
a wave mesoscopic system obeying  discrete resonances that also
appears in other areas of physics.
\end{abstract}

\maketitle


\noindent {\bf  Introduction}.  Concept   of mesoscopic systems most
often appears in condensed matter physics, e.g, in studying
properties of superconductors on a scale comparable with that of the
Cooper pairs\cite{condensed}, of miniaturized transistors on a
computer chip, of   disordered (glassy, granular) systems, when
self-averaging is inefficient  and fluctuations or the system
prehistory become important.  Similar situation occurs also in
various natural phenomena -- from wave turbulent systems in the
Ocean \cite{zak4} and Atmosphere, when wave lengthes are compatible
with the Earth radius~\cite{KPR}, to medicine \cite{med1}, and even
in sociology and economics, when finite size of a system
(population, sociological group, market) becomes important
\cite{opinion}. Mesoscopic regimes are at the frontier between
detailed, dynamical and statistical, self-averaging description of
systems. Important observation for finite-size, weakly-decaying wave
systems was made in \cite{K94}: discrete spacial-time resonances
form small isolated clusters of interacting modes without energy
exchange between the clusters. Clearly, there is exists relatively
short, ``threshold" wave, involving into  the cluster with  size
large enough to ``penetrate" into region of very short  waves, where
statistical description, that ignores the resonance discreetness, is
valid.

Mathematical problem of finding these clusters in concrete cases is
equivalent to solving some systems of high ($\gtrsim 12$) order
Diophantine equations on a space of 6-8 variables in big
integers~\cite{K98}. Recently developed algorithms for their
analysis~\cite{KK06-12} allows to find, in particular, all resonance
clusters  of atmospheric planetary waves, described by the spherical
functions $Y_\ell^m$ with eigenvalues $|m| \le\ell \le 50$. It
turned out that in this domain, consisting of 2500 spherical
eigen-modes $Y_\ell^m$, there exist only 20 different clusters
involving only 103 different modes. Moreover, 15 of these clusters
have the simplest ``triad" structure, formed by three modes.
Importantly, there are only four isolated triads in the domain
$0<m,\ell \le 21$, which is meteorological significant for the
problem of climate variability on intra-seasonal scale of about
10-100 days (waves with $\ell>21$ have to short period to play
significant role in this problem).

The main physical message of our Letter is that so-called
\emph{Intra-seasonal Oscillations (IOs) of the Earth atmospheric
flow can be rationalized as periodical energy exchange within the
above mentioned four isolated triads of the planetary waves}. IOs
has been firstly have been detected~\cite{mj1}
 in the study of   time series of tropical wind.
  Similar processes have been also discovered in the
atmospheric angular momentum, atmospheric pressure, etc. Detailed
analysis of the current state of the problem is presented in
\cite{gh3} and references therein; most part of the papers are
devoted to the detection of these processes in some data
sets,~\cite{all3,all1} and  to the reproducing them in computer
simulations with comprehensive numerical models of the atmosphere
~\cite{all}.  Nevertheless,  many aspects of the IOs remain
unexplained, \emph{e.g.} the reason of IOs in the North Hemisphere
is supposed to be topography, see e.g.~\cite{all2} and no reason
is given for IOs in the South Hemisphere, there is no known way to
 predict the appearance of IOs, etc.

 Our model considers IOs  as intrinsic  atmospheric phenomenon,
related to  a system of resonantly interacting triads of planetary
waves, which is an example of wave mesoscopic system. The model is
equally applied to the North and  South Hemispheres, is independent
(in the leading order) of the Earth topography,   naturally have the
period of desired order and allows to interpret the main observable
features of IOs (see Sec.~5).

\noindent {\bf 1.  Atmospheric planetary waves}. 
These waves are classically studied in the frame of barotropic
vorticity equation on a sphere:
 ~\cite{KPR}:
\be\label{BWE}
 \partial \triangle \psi\big/ \partial t  + 2\,  \partial
\psi\big/ \partial \lambda  +   J(\psi,\triangle \psi) =0\ .\ee
 Here
$\psi$ is the dimensionless stream-function; velocity $\B v=
\Omega\,  R  \, [\bf   z  \times \bm \nabla \psi] $, with $\Omega$
being the angular velocity   of the  Earth and  $\bf z$ -- the
vertical unit vector;  the variables $t, \varphi$ and $\lambda$
denote dimensionless time (in the units of $1/\Omega$), latitude
($-\pi/2 \leq \varphi \leq \pi/2$) and longitude ($0 \leq \lambda
\leq 2\pi$) respectively; $\triangle \psi$ and $J(a,b)$ are
spherical Laplacian and Jacobian operators.
  The linear part of this
equation has solutions in the form   $
 A_j \, Y_{\ell_j}^{m_j} (\lambda,\varphi) \exp(i\omega_j  t)$, $
\omega_j=-2m_j/\ell_j(\ell_j+1)$. Integer parameters $\ell_j$ and
$(\ell_j-m_j)\geq 0$ are   longitudinal and latitudinal wave-numbers
of $j$-mode, they are equal to the number of zeros of  the spherical
function along the longitude and latitude.

Assuming small level of nonlinearity, $|A_j|\ll 1$,  we restrict
ourselves  by resonant interactions only. Under the resonance
conditions for three modes: $\displaystyle
\omega_1+\omega_2=\omega_3 \,, \quad m_1+m_2=m_3 \,, $   in which  $
|\ell_1-\ell_2| < \ell_3 < \ell_1+\ell_2,$ and $
\ell_1+\ell_2+\ell_3$  is odd, the triad amplitudes $A_j(t)$ varies
in time according to equations~\cite{KPR}:
 \bea \nn  N_1
d{A}_1/ dt &=&  2\,  Z N_{32}\, A_3A_2^*, \  N_2 \,  d{A}_2/ dt= 2\,
Z\, N_{13} \,A_1^*A_3, \\ \label{complexA} N_3 d{A}_3/ dt&= & 2\, Z
N_{21} \,A_1A_2 \,, \quad \mbox{for} \ |A_j|\ll 1\ .
 \eea
Here $N_j\=\ell_j(\ell_j+1)\,, \ N_{ij}\= N_i-N_j$,   and
interaction coefficient $Z$ is an explicit function of wave numbers.
This system
 conserves energy $E$ and enstrophy $H$: 
\be  \label{ints} E =  E_1+E_2+E_3\,,\quad  H = N_1E_1 + N_2E_2 +N_
3E_3\,, \ee 
where the energy of $j$-mode is $ E_j\= N_j  |A_j |^2$.

\begin{figure}\begin{center}
\vskip -0.2cm
 \includegraphics[width=9cm,height=5cm]{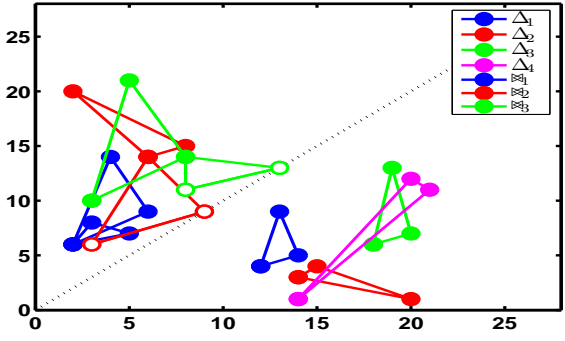}
 \end{center}\vskip - 0.8cm
\caption{\label{f:1}Four isolated triads are shown   below the
diagonal dash line;   $\ell$ -- along  horizontal axis and $m<\ell$
along the vertical one. The three clusters of two connected triads
are shown   above diagonal dash line (now $\ell$ -- along  the
vertical axis and $m\le\ell$ -- along the horizontal one). The
(meteorological  significant) spectral domain is restricted by $0<m,
\, \ell \le 21$. Localized modes are shown by full circles,
delocalized ones (with relatively small interaction amplitude) by
empty circles. }
\end{figure}


{\bf  2. Classification  of the triads}. Consider  the structure and
properties of interacting resonant triads in the meteorological
significant domain  $0<m, \, \ell \le 21$, where we found  four
isolated triads, denoted as $\Delta_1,\dots\,,\Delta_4$, three
"butterflies", i.e. clusters of two triads   (denoted as $\bowtie_{
1}$, $\bowtie_{ 2}$ and $\bowtie_{3}$) that are connected by a
common mode, and one cluster of 6 connected triads denotes as
$\boxtimes $. The structure of all isolated resonant triads and
"butterflies" clusters is shown in Fig.~\ref{f:1}.  Main information
about the triads in the chosen spectral domain  is given in  left 4
columns on   Table 1: notations of the triads, three pair of
$\ell_j, m_j$ for each triad, the value of the interaction
coefficient $Z$ and the so-called ``interaction latitude"
$\varphi_0$ introduced in Ref.~\cite{KPR}. Columns 5-7 contain data
which is necessary to compute period of resonant interactions and
will be commented on further.

We can interpret the latitude $\varphi_0$ as follows. The overlap of
three wave-function in a triad, $ \C Z ( \lambda, \varphi )\=
Y_{\ell_1 }^{ m_1}( \lambda, \varphi ) Y_{\ell_2 }^{ m_2}( \lambda,
\varphi )Y_{\ell_2 }^{ m_3}( \lambda, \varphi )$  shows a
contribution to the interaction coefficient $Z\propto \int \C Z (
\lambda, \varphi )\, d \lambda d \varphi$ from a particular location
on the sphere. The overlap  $ \C Z ( \lambda, \varphi )$

\begin{tabular}{|c|c|c| c||c| c|c|c|}
  \hline 
   Triad & Modes $[m,\ell]$ &  $Z$ & $\varphi_0$  & $K[\mu]$ &
$10^{6} E_0
$ & $T_0$\\
\hline
$\Delta_1$ & [4,12] [5,14] [9,13] & 7.82 & 34  & 1.62 & $14.4$ &  $24$\\

\hline
$\Delta_2$ & [3,14] [1,20] [4,15] & 37.46 &19    & 1.14 & $5.4$ &  $5$\\
\hline
$\Delta_3$ & [6,18] [7,20] [13,19]  & 13.66 &34   & 1.74 & $32.0$ & $19$\\
\hline
$\Delta_4$ & [1,14][11,21][12,20] & 47.67 & 28    & 1.21 & $0.58$  & $13$\\
\hline \hline
$\bowtie_{1 }$ & [2,6] [3,8] [5,7] &3.14 &  35    & 1.64 & $5.08$ &  $30$\\
  & [2,6] [4,14] [6,9] & 14.63 & 37     & 1.61 &
$0.395$ &   $10$\\
\hline $\bowtie_{2 }$ & [6,14] [2,20] [8,15] &  69.25 & 31  & 1.13 &
$0.61$ & $8$ \\
  & [3,6] [6,14] [9,9] & 11.31 &$-$   & 1.17  & $0.360$ & $13$ \\
\hline $\bowtie_{3 }$ & [3,10] [5,21] [8,14] & 61.99 & 31  & 1.27 &
$0.133$ & $7$ \\
 & [8,11] [5,21] [13,13] &8.71 &$-$   & 1.36 & $0.784$ & $24$\\
\hline \hline
$\boxtimes $ & [1,6] [2,14] [3,9] & 28.98 &17    &  1.38 & $0.247$ &   $6$ \\

  & [2,7] [11,20] [13,14] & 2.77 &42   & 1.08 & $1.78$  &  $26$\\

  & [1,6] [11,20] [12,15] &15.08 & 29    & 1.06 &
$0.262$ &  $11$\\

  & [9,14] [3,20] [12,15] &74.93 & 50   & 1.36 & $0.487$ &  $8$ \\

 & [3,9] [8,20] [11,14] & 32.12 &40   & 1.11 & $0.251$ &  $9$ \\

  & [2,14][17,20][19,19] & 11.05 &$-$  &  1.05 & $3.33$ & $24$ \\
\hline
\end{tabular}~\\

\noindent {\small Table 1. For each triad  the following data  are
given: all resonantly interacting modes, interaction coefficient
$Z$, interaction latitude $\varphi_0$ (in grad), magnitude of the
elliptic
 integral $K(\mu)$, corresponding to the ESMRW
   December 1989 data for 500 hPa  initial energy distribution
in a triad, and initial dimensionless energy $E_0*10^6$ of each
triad and resulting   $T_0$ values (in days).}\\~\\
\noindent
has a maximum at a particular latitude $\varphi_0$ and a
narrow latitudinal belt around $\varphi_0$ gives  the main
contribution to the global interaction amplitude $Z$. That is why
$\varphi_0$ can be understood as the interaction latitude.

{\bf  3. General solution of the triad equations}. Linear change of
variables $B_i=\a_i A_i$ with $\a_i$ being explicit functions on
$N_j$ allows to rewrite Sys.~(\ref{complexA}) as \\
 $ \displaystyle ~~~\dot{B}_1=  2 Z B_2^*B_3,\quad
\dot{B}_2=  2 Z B_1^* B_3, \quad \dot{B}_3=  2 Z B_1 B_2\ .$\\
This system  has two
 independent conservation laws

\begin{figure*}
\begin{center}\vskip -0.2cm
\includegraphics[width=6.1cm]{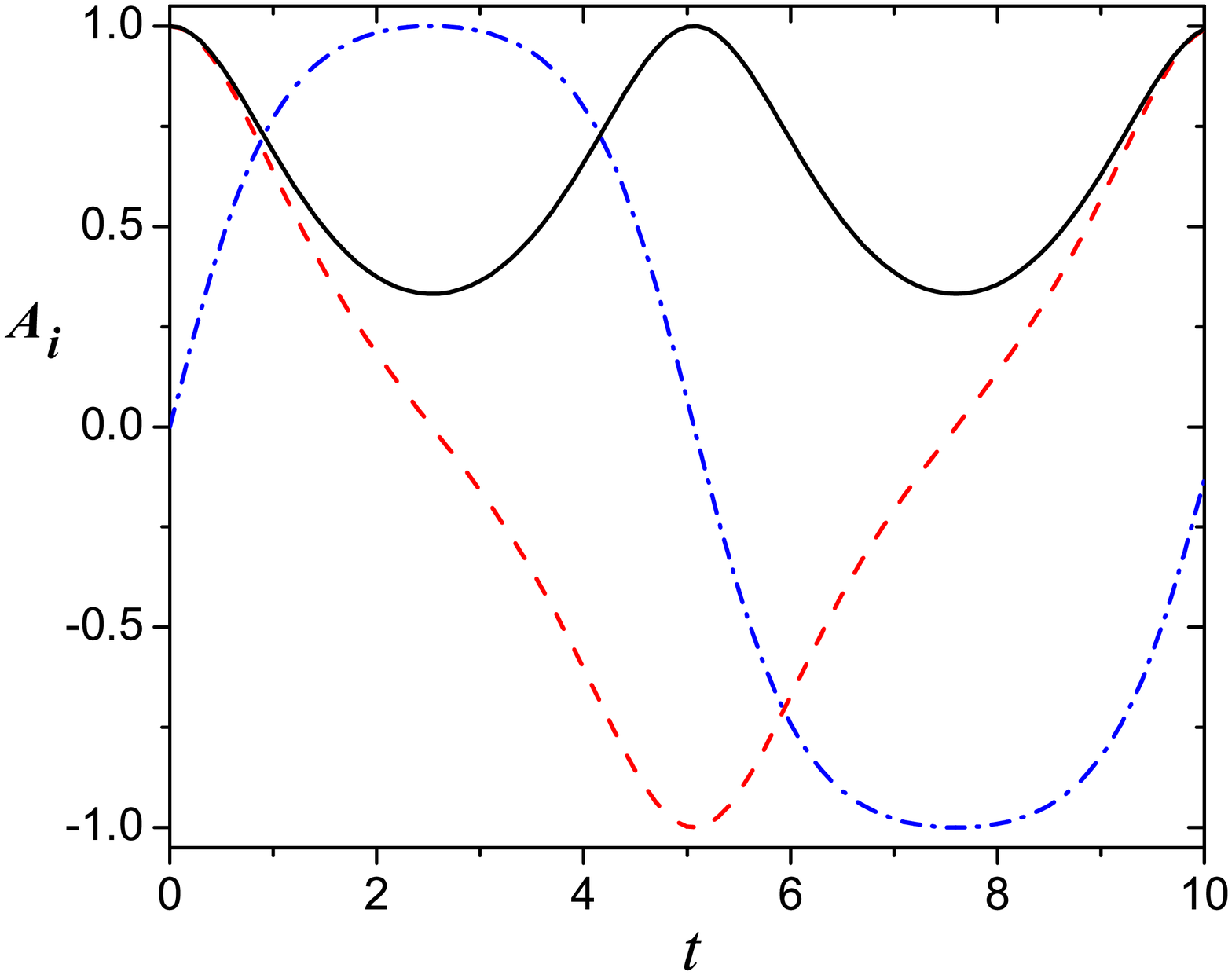}
 \includegraphics[width=5.8  cm]{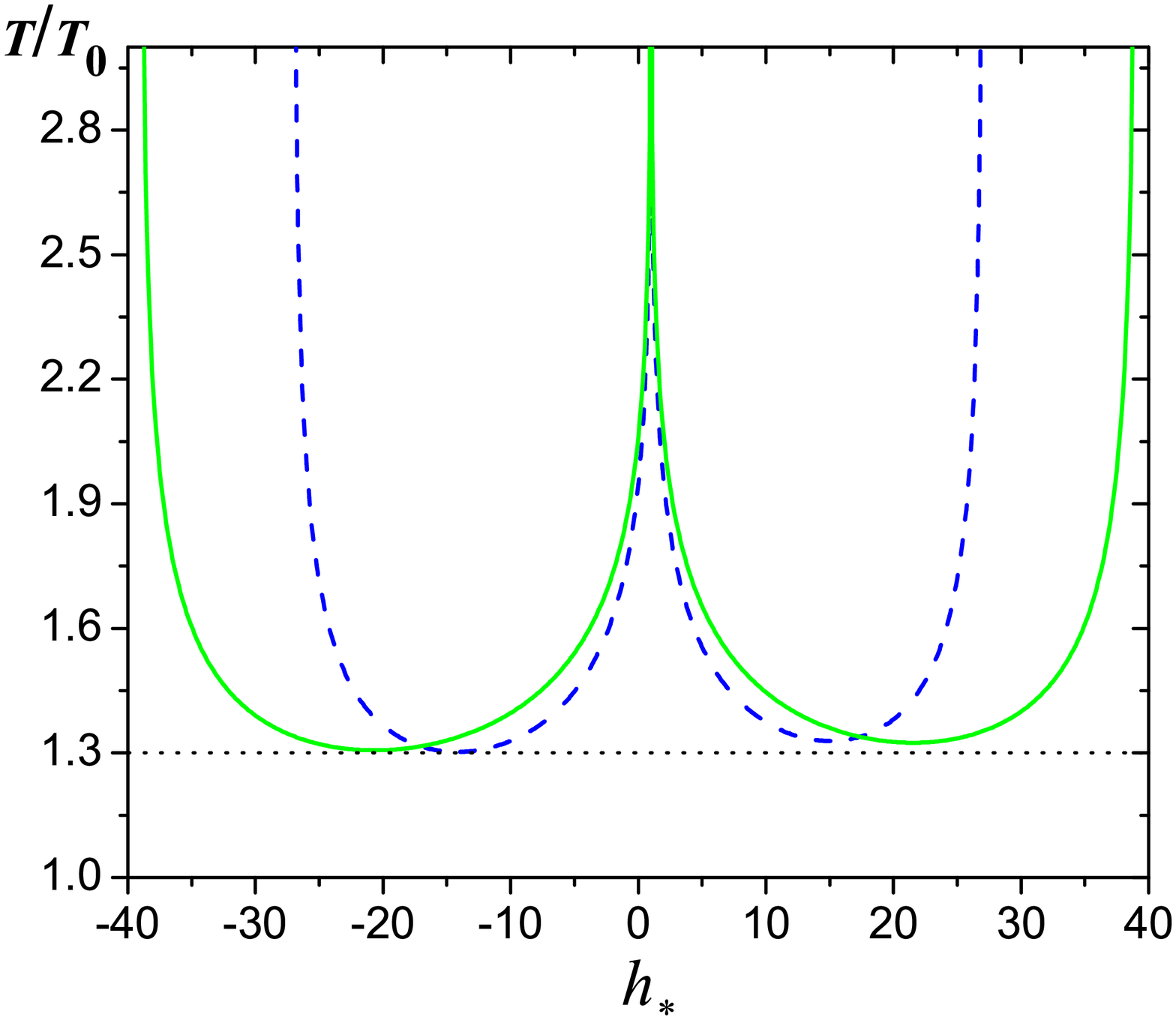}
  \includegraphics[width=5.8 cm]{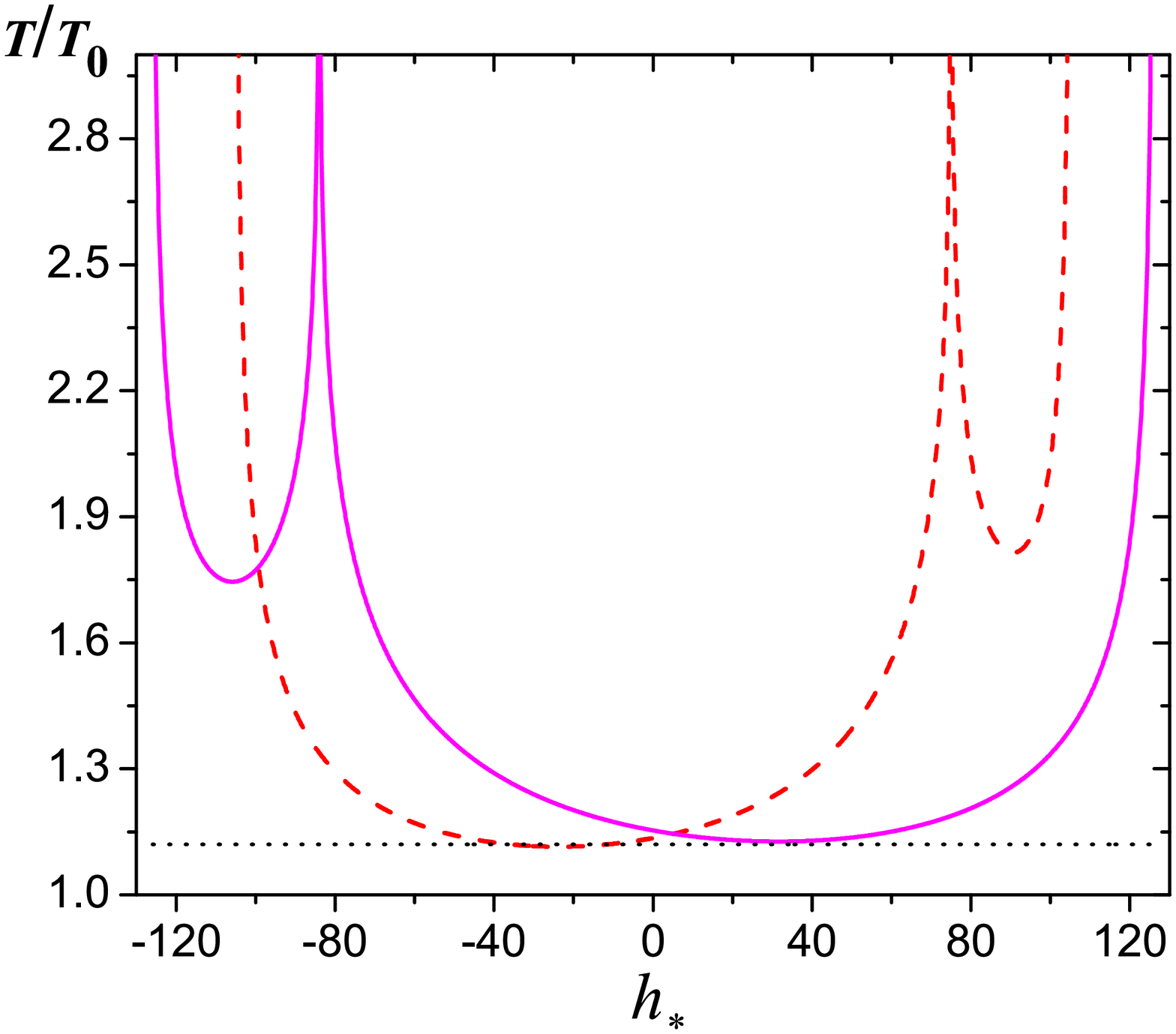}
\end{center}\vskip -0.6cm
\caption{\label{f:2} Left panel: Time dependence of  $A_1$, $A_2$
and $A_3$ [denoted by green solid,  red dashed and blue dot-dashed
lines accordingly] of the triad $\Delta_1$, with $\mu=0.89$,
corresponding to observed data.  Middle and Right panels: Dependence
of the triad periods $T(E,h)/T_0(E)$, \REF{tauha}.  Middle panel:
blue-dashed line corresponds to $\Delta_1$ triad, green-solid line
-- $\Delta_3$ triad. Right panel: Red dashed line -- $\Delta_2$ and
magenta solid line -- $\Delta_4$ triad. Black dashed lines denote
minimal period: $T\approx 1.3 T_0$ for $\Delta_1$, $\Delta_3$ and
$T\approx 1.18 T_0$ for $\Delta_2$, $\Delta_4$.}
\end{figure*}

 \BSE\label{MR} \bea \label{MRa}
 I_1&=&|B_2 |^2 + |B_3|^2 =( E\, N_1-H ){N_{23} }/{N_1 N_2
 N_3}\,,~~~~\\
 I_2&=& |B_1 |^2 + |B_3|^2 =( E\, N_2-H ) {N_{13 }}/{N_1 N_2
 N_3} ~~~~\,,
\eea\ESE
which are linear combinations of the energy $E$ and enstrophy $H$.
Direct calculations show that the general solution for   $B_i$ is
expressed in Jacobian  elliptic functions 
 $
B_1 = B_{1,0} {\rm  cn}\, (\tau  -\tau_0)$, $B_2=B_{2,0}{\rm sn}\,
(\tau -\tau_0)$, $B_3=B_{3,0} {\rm  dn}\, (\tau -\tau_0)$,
 where
$B_{j,0} \,, \ \ \tau_0 \,, $ are defined by initial conditions
and $\tau =t / 2\, Z \sqrt[4]{I_1I_2}\ . $ Functions  cn$(\tau)$,
sn$(\tau)$ and dn$(\tau)$ are periodic with the period $4K(\mu),
4K(\mu)$ and $2K(\mu)$ correspondingly, were $K(\mu)$
$$
 K(\mu)= \frac 2 \pi \int _0 ^{\pi/2} \frac{ d \theta}{ \sqrt{1-\mu
\sin^2\theta}} \,, \ \mu^2\equiv \min \Big\{ \frac{I_1}{I_2}\,,
 \frac{I_2}{I_1}  \Big\}  \le 1 \ . 
$$
Fig~\ref{f:2} illustrates the typical time dependence   of all
three dimensionless amplitudes of the triad $\Delta_1$.  One sees
that $K(\mu)$ is a smooth function that changes slowly enough such
for the wide region of the initial conditions it can be roughly
considered as a constant.

\noindent {\bf  4. Period of triad oscillations}. The period of
energy exchange (measured in days) in the triads  is given by
$\label{tauX}T = {\pi \, K(\mu)} \Big / { Z \sqrt[4]{I_1I_2}}$, that
can be written a product of functions $E$ and   the  \emph{ ratio of
the enstrophy to the energy} $h\= { H}\big /{ E}$ :
\BSE\label{tauh}
 \bea 
\label{tauha}T&\RA&  T(E,h)  =  T_0(E)\,   K (\mu) \, f(h)\,, \\
T_0(E)&=&\frac{\pi}{2\, Z \sqrt{2 E}}\, \sqrt{\frac{N_1 N_2 N_3
}{N_{21} \sqrt
{N_{31} N_{23} } }}  \,,~~~~~\label{tauhb} \\
f(h) &=& \sqrt{ N_{12}\big/2 \sqrt {(N_1-h)(h-N_2)}} \
\label{tauhc}.
 \eea\ESE
Here $K$ depends on $\mu$ and, in its turn, $\mu$ depends on $h$ as
 ~~$\displaystyle
  \mu^2(h)  = \min \Big[ \ \frac{( h-N_2)
 N_{23} } {( N_1 - h) N_{31} }\,,
 \frac{ N_1 -
h) (N_{31} }{(h-N_2)
 N_{23} } \Big] $.
Equation~\Ref{ints} show that possible values   of $h$ lie inside
one of the two intervals $
 N_2 \ge h \ge N_1$   or $N_1 \ge h \ge N_2$.
Without loss of generality we set $
 N_2 \ge h \ge N_1 $,
then  maximal possible value $h=N_2$ is realized  if $E_1=E_3=0$,
i.e. only the second mode is excited.  The minimal value $h=N_1$ is
possible if $E_2=E_3=0$, i.e. only the first  mode is excited. In
both cases, according to basic Eqs.~\Ref{complexA} there is no time
evolution, i.e. $E_j=$ const. for $j=1,2$ and $3$. This is in
agreement with Eq.~\Ref{tauhc}, according to which $f(h)\to \infty$
for $h\to N_2$ or $h\to N_1$.

Function $f(h)$, \REF{tauhc}, has a minimum (equal to one) just in
the middle of the interval  $
 N_2 \ge h \ge N_1 $ at $
h=h_+\= (N_1+N_2)/2$.
 For this value of $h$
 $\displaystyle \hskip 2 cm
\mu^2(h_+)  = \min \Big\{ \ \frac{ (N_2-N_3) } {(N_3-N_1) }\,,
\frac{ (N_3-N_1) }{
 (N_2-N_3) } \Big\}
 $.
For isolated triads of interest $\Delta_1\,,\Delta_2\,, \Delta_3$
and  $ \Delta_4$\,,  the values of $\mu^2(h_+)$ are 0.93, 0.41, 0.97
and 0.45 respectively, with $K(\mu)$ equal to 1.96, 1.27, 2.22 and
1.30.

The less trivial case of infinite period corresponds to    $\mu=1$,
which is realized at $ h= h_{\rm crit}\=N_1+N_2-N_3$. In this case
 $B_3(\tau)= \tanh \tau $ and for $\tau\to \infty$, $B_3\to 1$ and
 $B_1$ and $B_2$ exponentially fast go to zero, i.e.
  for the
 specific value
  $h= h_{\rm crit}$, the  high mode $B_3$ exponentially fast
takes energy from two low modes. This is possible only for three
particular values of $h$:  $h=N_1$, $N_2$ and $N_1+N_2-N_3$. The $h$
dependence of the period of the   triads $\Delta_1\,, \dots\,,
\Delta_4$,  is
 presented in  Fig.~\ref{f:2}.  One sees, that the regions, where the
 period exceeds twice the minimal possible are very narrow, just
 few percent of the available interval of $h$.  This
 means, that though theoretically for {\em  each triad we can always
  choose initial conditions in such a way that period will be
  large and even tend to infinity}, the probability of this is
very small.

Indeed, for qualitative analysis we can think that in the turbulent
atmosphere the probability  to get some energy from global
disturbances  to a particular planetary wave is independent from the
state of other waves and this probability is more or less the same
for each wave in a triad. If so, the probability $\C P(h)$ to have
initial conditions with some value of $h$ has to be a smooth
function of $h$ in the whole available interval  $
 N_2 \ge h \ge N_1 $. Roughly speaking,
we can approximate  $\C P(h)$ as the constant: $
 \C P(h)\simeq  1 \big / ( N_2-N_1)$, $N_2 \ge h \ge N_1$.
With this approximation we can, for example, for triads
$\Delta_1$--$\Delta_4$  estimate
 the probability to have the period, twice
exceeding the minimal one $T_0(E)$, \REF{tauhb}, as few percents.
Moreover, as one sees in Fig.~2, the typical value of the period
$T$ is about $(1.4\pm 0.1)T_0$ for the triads $\Delta_1$,
$\Delta_3$ and about $(1.2\pm 0.1)T_0$ for the triads $\Delta_2$,
$\Delta_4$.  This conclusion is in a qualitative agrement with the
ECMWF (European Center for Medium-Range Weather Forcast) winter
data, shown in Table 1, column 7.

\noindent{\bf  5. \!\!Intra-seasonal \!\!Oscillations \!\!as
\!Resonant \!Triads.} Our interpretation of  IOs as dynamical
behavior of $\Delta_1\,,\Delta_2\,, \Delta_3$ and $\Delta_4\ $
triads allows one to answer
some questions appearing from meteorological observations \cite{gh3}.\\
\textbullet~{\it What is the cause of IOs in South Hemisphere?  }
 The basic fact of our model is the
existence of global nonlinear interactions among
 planetary waves, independent of the topography. \\
\textbullet~{\it Why the period of so-called
 "topographic" oscillation in North
Hemisphere is given as 40 days by some researchers and 20-30 days -
by other researchers?  } The  variations in  the magnitudes of the
period are caused by different initial energy and/or initial energy
distribution among
the modes of the same triad.  \\
\textbullet~{\it How do the tropical and mid-latitude oscillations
interact?  } Two mechanisms are possible: i) Triads with
substantially different interaction latitudes belonging to the same
group, for instance, triads [(1,6) (2,14) (3,9)] and [(3,9) (8,20)
(11,14)] of $\boxtimes$   exchange their energies through other
modes of this group and belong correspondingly to the tropical and
extra-tropical latitudinal belts, and ii) Isolated triads can
interact
 via some special modes  called {\it active
near-resonant modes} \cite{K94}. These modes have  smallest
resonance width
 with a given triad, and  are themselves parts of
some other resonant triad. For instance, the mode (13,19) is a near
resonant for $\ \Delta_4\ $ (with resonance discrepancy
$\delta=0.16$) and is resonant for $\ \Delta_3$.\\ 
\textbullet~{\it Why do the intra-seasonal oscillations are better
observable in winter data?  } In summer, modes have  higher
energies, periods of the triads become smaller, and resonances with
big enough resonance width can destroy the clusters. 
\\
\textbullet~{\it How to predict these recurrent features?}
Amplitudes of the spherical harmonics with wave numbers taken from
Table 1 have to be correlated: $\< A_1(t)A_2(t)A_3^*(t)\> \sim
\sqrt{\< |A_1(t)|^2\>\< |A_2(t)|^2 \>\<|A_3(t)|^2\>}$, see Fig. 2.
Magnitudes of the expected periods can be computed beforehand by the
given explicit formulae.

\noindent {\bf  6. Conclusions}

\textbullet~ Our simple model provides main robust features of IOs
in terms of resonance clusters consisting of three modes of
atmospheric waves.

\textbullet~  Energy behavior within the bigger clusters should be
a subject of a special detailed study. Knowledge of cluster
structure allows to simplify drastically their analysis. For
instance, for "butterfly" cluster at least 6 real integrals of
motions  can be easily found. Universal method to construct
isolated clusters and write out explicitly corresponding dynamical
equations for a wide class  of mesoscopic systems is given in
\cite{KM07}.

\textbullet~ Our approach is quite general and can be used for
studying many other mesoscopic systems, provided that explicit
form of dispersion function $\ \o(\B k) \ $ is known (here $\B k$
is the wave vector of plane systems with periodical boundary
conditions, or another set of eigenvalues in more complicated
cases, like $\ m, \ell \ $ for the sphere).  Properties of a
specific mesoscopic system will depend on the 1) form of $\ \o(\B
k)$,  2) dimension of $\B k,$ 3) number of conservation laws,  4)
initial magnitudes of the conserved values (energy, enstrophy,
etc.) and their initial distribution among the modes in the
cluster.

\noindent {\bf Acknowledgement}.  We  express our gratitude to
Vladimir Zeitlin, anonymous referees and specially to Grisha Volovik
for various comments and advises.  We are thankful to Yuri Paskover,
Oleksii Rudenko and
 Mark Vilensky for  stimulating discussions and help. We also
acknowledge the support of the Austrian Science Foundation (FWF)
under projects SFB F013/F1301,F1304 and of the US-Israel Binational
Science Foundation.

\end{document}